\documentstyle[twoside,fleqn,espcrc2,epsfig]{article}
\newcommand{\AmS}{{\protect\the\textfont2
  A\kern-.1667em\lower.5ex\hbox{M}\kern-.125emS}}

\title{Neutrino Mass Difference Induced Oscillations in Observed Muon Decays}
\author{
Y.N. Srivastava 
\address{Physics Department \& INFN, University of Perugia, Perugia, Italy},
S. Palit and A. Widom   
\address{Physics Department, Northeastern University, Boston MA 02115 U.S.A.},
E. Sassaroli 
\address{Laboratory for Nuclear Science, MIT, Cambridge MA U.S.A.}
}

\begin{document}

\begin{abstract}
The recent real time KARMEN anomaly for the electron neutrino counting 
rates (obtained from stopped muon decays) is analyzed by employing a 
neutrino flavor rotation model in which the muon neutrino is in a 
superposition of only two mass eigenstates. On the basis of experimental  
electron neutrino counting oscillations, we find a neutrino mass splitting 
$|m_2^2-m_1^2|=(0.22\pm 0.02)(eV/c^2)^2$ and a flavor rotation angle 
$\phi =0.34\pm 0.10 $, both within a $95\%$ confidence interval.
  
\end{abstract}
\maketitle

Recent {\em real time observations} of the electron neutrino $\nu_e $, 
produced from a $\mu^+$ decay 
\begin{equation}
\mu^+\to e^+ +\bar{\nu }_\mu+\nu_e ,
\end{equation}
indicate an oscillatory counting rate\cite{1}. These data provide 
experimental evidence for a neutrino mass matrix with flavor rotations. 
The oscillations of interest are made manifest in the experimental KARMEN 
anomaly\cite{1}. Our purpose is to discuss this fact, employing a model 
in which the muon neutrino is in a superposition of two mass eigenstates 
\begin{equation}
<\bar{\nu}_\mu |=\cos \phi <\bar{\nu}_1|+
\sin \phi <\bar{\nu}_2|,
\end{equation}
with mass eigenvalues of $m_1$ and $m_2$, respectively. 

One may view Eq.(1) as occurring in two stages: (i) 
\begin{equation}
\mu^+\to W_{eff}^+ +\bar{\nu }_\mu,
\end{equation}
and (ii)
\begin{equation}
W_{eff}^+ \to e^+ +\nu_e. 
\end{equation}
The virtual ``effective'' $W_{eff}^+ $ standard electroweak charged  
Boson is far off the $M_W$ mass shell, and appears in the lowest order 
tree diagram for Eq.(1) via the charged Boson propagator 
${\cal D}_{\alpha \beta }$. The effective wave function is given by 
\begin{equation}
W^+_{eff,\alpha }(x)=\int {\cal D}_{\alpha \beta }(x-y)
{\cal J}^{+,\beta} (y)d^4y,
\end{equation}
where the charged current  ${\cal J}^+$ is the source of the $W_{eff}^+$ 
Boson. In the far off $M_W$ mass shell regime of Eq.(1), it is 
sufficiently accurate to write the wave function in Eq.(5) using 
the field-current identity of the Fermi formulation 
\begin{equation}
W^{+,\alpha }_{eff}=g_W\Big({\hbar \over M_W c}\Big)^2
\bar{\Psi } \gamma^\alpha (1-\gamma_5)N_\mu ,
\end{equation}
Eq.(6) employs units for which the $W$ fine structure constant 
$\alpha_W=(g_W^2/\hbar c)$ is related to the Fermi coupling 
fine structure constant by $(\sqrt{2})\alpha_W=(G_F M_W^2/\hbar c)$.

In Eq.(6), the muon wave function $\bar{\Psi }$ is the one particle 
to vacuum matrix element of the 
muon field operator $\bar{\psi }$. The muon wave function is 
\begin{equation}
\bar{\Psi }(x)=<0|\bar{\psi }(x)|\mu^+>.
\end{equation}
The muon neutrino wave function $N_\mu$ is the vacuum to 
one particle matrix element of the muon neutrino field operator, 
\begin{equation}
\nu_\mu (x)=\cos \phi \nu_1(x)+\sin \phi \nu_2(x).
\end{equation}
The neutrino wave function is   
\begin{equation}
N_\mu (x)=<\bar{\nu }_\mu|\nu_\mu (x)|0>.
\end{equation}
From Eqs.(2), (8) and (9), one finds that the muon neutrino wave function 
has the oscillation form
\begin{equation}
N_\mu (x)=\cos^2\phi N_1(x)+\sin^2 \phi N_2(x).
\end{equation}

Perhaps more surprising, but unambiguously true from Eqs.(6) and (10), 
is the notion that the decay Eq.(3) of a muon with fixed four 
momentum $P_\mu$ throws the other particle\cite{2}, \cite{3} 
(here the $W_{eff}^+$) 
{\em into an oscillation superposition of amplitudes }\cite{4}, \cite{5} 
\begin{equation}
W_{eff}^+(x)=\cos^2 \phi W_1^+(x)+\sin^2 \phi W_2^+(x).
\end{equation}
In Eq.(11), 
$W_{j,\alpha }^+\propto (\bar{\Psi }\gamma_\alpha (1-\gamma_5)N_j)$ 
for $j=1,2$.

The four momentum kinematics may now be written as follows: Let $P_\mu $ 
be the initial four momentum of the muon, and $p_j$ for $j=1,2$ the four 
momentum of the possible neutrino mass eigenstates. Then 
\begin{equation}
P_\mu^2=-(M_\mu c)^2,\ \ p_j^2=-(m_jc)^2,\ \ (j=1,2).
\end{equation}
Since total four momentum is conserved at each vertex of a Feynman 
diagram, even for internal virtual particles, one then associates two 
possible four momenta for the $W_{eff}^+$, 
\begin{equation}
P_{effW,j}=(P_\mu -p_j), \ \ (j=1,2)
\end{equation}
whose effective (virtual off shell) mass $M_{effW} $,  
\begin{equation}
(M_{effW}c)^2=-P_{effW,1}^2=-P_{effW,2}^2,
\end{equation}
is much smaller than the on shell mass of the $W^+$; i.e. 
$M_{eff}<<M_W$. From Eqs.(12), (13) and (14) it follows that 
\begin{equation}
2(p_2-p_1)\cdot P_\mu =(m_1^2-m_2^2)c^2.
\end{equation}
In the rest frame of the muon, the two possible neutrino energies,  
$\epsilon_1$ and $\epsilon_2$ have a Bohr transition frequency 
$\omega $ determined by Eq.(15); It is
\begin{equation}
\hbar \omega=\big|\epsilon_2-\epsilon_1\big|=
\Big({|m_2^2-m_1^2|\over 2M_\mu }\Big)c^2.
\end{equation}

The central result of this work now follows from Eq.(6) for 
the virtual $W^+_{eff}$ wave function. This wave function contains a 
product of a neutrino wave function $N_\mu (x)$ and a muon wave 
function $\bar{\Psi}(x)$. After absolute value squaring the $W^+_{eff}$ 
decay amplitude, one finds (from the neutrino wave function) the usual 
neutrino oscillation probability 
$$
\Big|\cos^2 \phi e^{-i\epsilon_1t/\hbar}+
\sin^2 \phi e^{-i\epsilon_2t/\hbar} \Big|^2=
$$
\begin{equation}
\big(1-\sin^2(2\phi )sin^2(\omega t/2)\big),
\end{equation}
and (from the muon wave function) the usual muon decay probability 
$\exp(-\Gamma t)$. Thus, for the probability $P(t)$ of the   
{\em $W^+_{eff}$ decay survival}, we have the differential equation     
\begin{equation}
-{dP(t)\over dt}=Z\Gamma e^{-\Gamma t}
\big(1-\sin^2(2\phi )sin^2(\omega t/2)\big).
\end{equation}
In Eq.(18), the constant $Z$ is found by normalizing the the total 
$W^+_{eff}$ survival probability; i.e. 
\begin{equation} 
-\int_0^\infty {dP\over dt}dt=P(0)-P(\infty )=P(0)=1.
\end{equation}

To test the above notion of observing oscillations in the muon 
neutrino $\bar{\nu }_\mu $, we analyzed the KARMEN anomaly data 
using our central Eq.(18). In the KARMEN experiment
\cite{6}, \cite{7}, \cite{8}, the electron 
neutrinos from $\mu^+\to e^+ +\bar{\nu }_\mu+\nu_e $ were detected 
at times $\{t_i \}$ (after the muons were stopped) within time bins 
of width $\Delta t_i =0.5\mu s$. 
 
One expects (using our theory) a mean number of detected 
electron neutrinos (from the muon decays) in each bin to be given by  
\begin{equation}
\mu_i(\omega ,\phi )=
-{\cal N}\Big({dP(t_i;\omega ,\phi )\over dt_i}\Big)\Delta t_i ,
\end{equation}
where ${\cal N}$ is chosen so that $\sum_i\mu_i$ is the (total) number 
of observed $\nu_e$ events. If the events in each time bin obey 
Poisson statistics, then the probability distribution for a counting 
sequence of $n_i$ events in the $i^{th}$ bin is given by 
\begin{equation}
{\cal P}[n;\omega ,\phi ]=\prod_i \Big({e^{-\mu_i(\omega ,\phi )}
\mu_i(\omega ,\phi )^{n_i} \over n_i!}\Big).
\end{equation}
The likelihood function (for the theoretical parameters $\omega $ and 
$\phi $) is then determined by experimental KARMEN data $n_{i,data}$ 
via 
\begin{equation} 
\Lambda (\omega ,\phi )={\cal K}{\cal P}[n_{data};\omega ,\phi ],
\end{equation}
where ${\cal K}$ is an arbitrary constant. Under the best circumstances, 
the likelihood function exhibits a single maximum in the theoretical 
parameter space $(\omega ,\phi )$.

We employed this method of determining the theoretical parameters 
$\omega $ and $\phi $, thus analyzing the KARMEN data from the viewpoint 
of the $\bar{\nu }_\mu$ induced oscillating $W^+_{eff}$ decays. The 
likelihood function is exhibited in Fig.1. 
\begin{figure}[htbp]
\begin{center}
\mbox{\epsfig{file=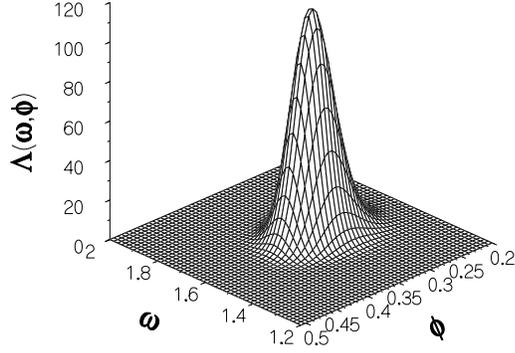,height=70mm}}
\caption{The likelihood as a function of the parameters 
$\omega $ and $\phi $.}
\label{kfig1}
\end{center}
\end{figure}
A single maximum 
likelihood peak was found. The derived values of $\omega$ 
and $\phi$ were determined to be  
\begin{equation}
\omega =(1.60\pm 0.16)/\mu s , 
\end{equation}
and
\begin{equation}
\phi =0.34\pm 0.10 ,
\end{equation}
within a $95\%$ confidence interval obtained from the likelihood function. 
In terms of the neutrino mass squared differences, 
\begin{equation}
\big|m_2^2-m_1^2\big|=(0.22\pm 0.02)\big(eV/c^2\big)^2 ,
\end{equation}
also within a $95\%$ confidence interval.

In Fig.2, 
\begin{figure}[htbp]
\begin{center}
\mbox{\epsfig{file=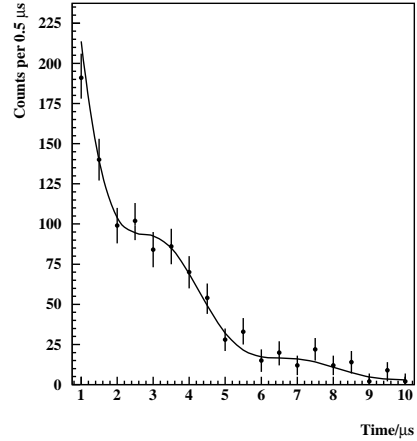,height=70mm}}
\caption{The theoretical time dependence of the $\nu_e$ counts along with 
the experimental points. }
\label{kfig1}
\end{center}
\end{figure}
we compare the theoretical real time oscillation 
in the $\nu_e$ counting rate (from the $\mu^+$ decays) with 
the {\em experimental} oscillation for the $\nu_e$ counting rate 
in each time bin. The theoretical parameters $(\omega,\phi)$ were 
chosen from the maximum of the likelihood function, The over 
all fit may be described by the statistical $(\chi^2 /dof)\approx 1.4$. 

In summary, the real time KARMEN anomaly for the electron neutrino counting 
rates (obtained from muon decay) has been analyzed. The data provides 
evidence in favor of a neutrino flavor rotation model in which the muon 
neutrino is in a superposition of two mass eigenstates. The neutrino mass 
splitting fits the data employing Eq.(25), and the flavor rotation angle 
fits the data employing Eq.(24). The theoretical fit to the 
experimental KARMEN anomaly appears quite satisfactory.

\end{document}